\documentclass[11pt]{article}

\def\jou#1#2#3#4{{#1} {\bf #2} (#4) #3}

\def\NPA{{Nucl. Phys.} A}
\def\NPB{{Nucl. Phys.} B}
\def\PLB{{Phys. Lett.}  B}

\def\PRL{Phys. Rev. Lett.}

\def\PRD{{Phys. Rev.} D}

\def\EPJC{{Eur. Phys. J.} C}

\def\c{\chi}
\def\p{\pi}

\def\lqcd{\Lambda_{\rm {QCD}}}

\begin{document}

\title{
Filling up the Two-Body Partial Decay Width of P-Wave Charmonia 
with the Color Octet Contribution
} 

\author{S.M.H. Wong\thanks{The author thanks the organizers for
their hard work towards this conference. This work was supported by 
the U.S. Department of Energy under grant no. DE-FG02-87ER40328.} \\ \\
{\it School of Physics and Astronomy, University of Minnesota,} \\
{\it Minneapolis, Minnesota 55455, U.S.A.}} 

\date{}
\maketitle

Quarkonia are special hadronic systems in which the constituent 
quarks because of their large mass move very slowly therefore they
are open to non-relativistic treatment. So quarkonium production
or decay, once supplemented with non-perturbative wavefunctions 
or matrix elements can be calculated using the small velocity 
as an expansion parameter \cite{bbl}. Recently some exclusive $\c_J$ 
decay channels have been re-measured at the Beijing Spectrometer 
by the BES collaboration \cite{bes}. An attempt to compare with 
existing calculations of the channels $\c \longrightarrow \p \p$
and $\c \longrightarrow p \bar p$ revealed that these were 
out-of-date. Various parameters have changed since the last
serious attempts at calculating the partial widths and the 
understanding on some of the hadronic wavefunctions has also improved. 
It is therefore necessary to re-examine the decay calculations. 

The decay processes of interest are best calculated using the
perturbative QCD based scheme of Brodsky and Lepage (BL) \cite{bl}.
In this scheme, decay amplitudes are given by a convolution of 
non-perturbative hadronic distribution amplitudes and the associated
perturbative hard part through which the hard momentum flows. 
Using this scheme with the latest parameters such as $\lqcd$, 
decay constants and distribution amplitudes of pion and nucleon,
it was found that in both channels the theoretical width of the
$c\bar c$ system fell far short of the experimental results. 
The widths for the decay into $\p\p$ are at most 23\% for $\c_0$ and 
28\% for $\c_2$ of the experimental data. Those for the decay into 
$p\bar p$ are even worse at only 8.3\% for $\c_1$ and 10.4\% for 
$\c_2$. The next logical step to find out why there are such large 
discrepancies is to try the calculations again using 
the improved BL scheme of Sterman et al \cite{bls} hoping 
that the advantages in the improved scheme could compensate
for whatever in the original scheme that might have caused the
deficiencies \cite{kw}. Unfortunately this is not to be. The new 
scheme actually yielded widths of the same magnitude as obtained
from the old. Varying all parameters within their uncertainties 
to enlarge the widths did not help either. The results were still
at least a factor of two or more below. The widths at the size of 
the experimental data remained unattainable. So the small 
theoretical results did not arise because of some artifact in 
the scheme or the poor choice of parameters used. 

While searching for an explanation, it became obvious that color
octet might be the answer as this was shown to be a necessary
component in inclusive P-wave charmonium decay to ensure infrared
finite answers \cite{bbl,bg}. Strangely color octet has never 
been considered within the context of exclusive reactions. 
At first sight it cannot be the answer because higher Fock 
state contributions are suppressed by large momentum flow within 
the BL scheme. On the other hand, a contradiction would arise
if none of the exclusive $\c_J$ decay channels, many-body decays 
included, requires the octet contribution. This is because the
octet state must be there to ensure infrared finite inclusive width, 
which is just the sum of partial widths of all exclusive channels. 

An examination of the charmonium wavefunctions show that in momentum
space the P-wave wavefunction is down by $k/M$ in comparison to
the S-wave due to angular momentum, where $k$ is the internal
momentum of the heavy quarks within the charmonium and $M$ is the
mass of the latter. Extending this consideration to the decay amplitude 
of the $\p \p$ and $p\bar p$ channels, it can be shown that by examining 
the large mass $M$ dependence within the BL scheme that the singlet
and octet amplitudes for the decay into $\p \p$ go like $1/M^3$
and those for $p\bar p$ channel both also have the same dependency 
of $1/M^5$ \cite{wo}. So despite the fact that the octet is 
a higher state, it is not suppressed relative to the valence state. 
The suppression of the higher state in this case can no longer be
thought of as one when the P-wave valence state is brought down
to the same level by the orbital angular momentum. 
Once this theoretical hurdle has been overcome, it remains to work
out new ingredients to be used in the calculations. One is the
yet unknown wavefunction of the color octet state and the other
is more about the practicality of what to do with the constituent 
gluon. Neither of these was ever needed or encountered in the 
calculations of exclusive processes. Nevertheless the former
can be modeled and fit to data, and the latter argued to yield
a dominant contribution if the constituent gluon line ends on the 
hard part in the Feynman graphs \cite{bks}. With these details
sorted out, it was found that indeed the color octet contributions
were capable of bringing the theoretical widths up to the size of
experimental data \cite{wo,bks}. Thus confirming the theoretical 
expectations and resolving the problem of too small theoretical 
widths.

\end{document}